\documentstyle[12pt,titlepage]{article}
\input epsf.sty

\renewcommand{\theequation}{\thesection.\arabic{equation}}

\font\grande=cmr10 scaled \magstep4
\font\medio=cmr10 scaled \magstep2
\outer\def\beginsection#1\par{\medbreak\bigskip
      \message{#1}\leftline{\bf#1}\nobreak\medskip\vskip-\parskip
      \noindent}

\def\laq{\raise 0.4ex\hbox{$<$}\kern -0.8em\lower 0.62 
ex\hbox{$\sim$}}
\def\gaq{\raise 0.4ex\hbox{$>$}\kern -0.7em\lower 0.62 
ex\hbox{$\sim$}}

\begin{document}
\bibliographystyle {unsrt}
\titlepage
\begin{center}{\grande Resonant and Non-resonant Amplification }\\
\vspace{5mm} 
{\grande of Massless Gauge Fields}\\
\vspace{5mm}
{\grande during an Oscillating Dilaton Phase}\\

\vspace{10mm}
Massimo Giovannini\footnote{
e-mail: M.Giovannini@damtp.cam.ac.uk}\\
{\em DAMTP, Silver Street, Cambridge, CB3 9EW, United Kingdom}\\
\end{center}
\centerline{\medio  Abstract} 
The resonant and non-resonant amplification of electromagnetic
inhomogeneities is investigated within a phase of coherent dilaton
oscillations.
We show, in some explicit toy model, how the critical energy density
bound applied to the amplified fluctuations of the massless gauge
fields can constrain the dilaton mass.
\noindent

\vspace{50mm}
\centerline{{\sl Accepted for publication in {\bf Physical Review D}}}
\newpage

\renewcommand{\theequation}{1.\arabic{equation}}
\setcounter{equation}{0}
\section{Introduction}

The coherent oscillations of a massive scalar field minimally coupled
to gravity are a possible  source of many cosmological problems which
can result in a quite stringent set of bounds on the value of the
scalar mass.  The implications of the so-called ``Polonyi potential''
 were analysed in different contexts \cite{1,2,3,4}.

The energy density stored in the coherent oscillations red-shifts
faster than the surrounding relativistic plasma and its contribution
might become soon quite significant depending upon the initial
amplitude of the scalar field at the beginning  of the oscillating
phase.  If the initial amplitude is of the
order of the Planck mass (as it seems natural to require for
interactions of gravitational strength) the red-shifted energy density
of the oscillating field might have catastrophic consequences for 
the big-bang nucleosynthesis. 

At the nucleosynthesis curvature scale $H_N(t_N)$
the background temperature should be         
$T_N \sim O(1 Mev)$, implying a bound for the reheating temperature
associated with the scalar field decay and a lower bound for the
scalar field mass which should be heavier than $10^{4} Gev$. 
The decay of the coherent oscillations might have dangerous
consequences even before the nucleosynthesis scale because baryogenesis
forbids a too high entropy release during the decay of the scalar
field ($\Delta S \laq 10^{5}$) in order not to wash-out any
pre-existing baryon asymmetry \cite{5}, turning, ultimately, into a further
lower bound on the mass which should be larger than $10^{14} Gev$.
 A possible solution to the Polonyi problem could seem to fine-tune the
initial amplitude of the classical background oscillations to be much 
smaller than 1 (in Planck units).

In this case on top of the constraints on the classical scalar background
one should also be aware of other possible bounds coming from the
scalar field fluctuations whose non-relativistic modes might be
amplified, for instance, during some inflationary epoch preceeding the
coherent oscillations regime \cite{2}.

String moduli are coupled to Standard Model fields only through Planck
scale suppressed interactions and their effective potential is exactly
flat at small coupling  in the supersymmetric limit, but can become
curved thanks to non-perturbative effects or because supersymmetry
breaking. String moduli, then, also suffer 
 of the Polonyi problem \cite{5,12}.
Among the moduli fields a very special role is played by the 
dilaton background which
provides a unified value for the gravitational and gauge coupling
constant at the String scale $\lambda_{S}$ and which then couples not 
only to the Einstein-Hilbert term but also, for example, to the
abelian and non-abelian gauge fields \cite{5b} present in the theory.
Precision tests on the equivalence principle would seem to imply a
lower bound for the dilaton mass $m$ \cite{5c} ($m> 10^{-4}~ev$) and,
unless we do  adopt the approach of \cite{5d} (where a very light
dilaton mass is postulated), the problem of the
coherent oscillations persists.

One possibility would be to fine-tune the initial amplitude of the dilaton
to be much smaller than one (in Planck units) and then
 the possible constraints coming from the
amplified dilaton fluctuations seem quite alleviated provided they are
amplified with spectrum increasing in frequency \cite{10} as it seems
reasonable in a string cosmological scenario \cite{11}. 
 There have been recently few interesting  suggestions in order to solve the
moduli problem . A second stage of inflation able to dilute
the energy of the moduli field was suggested \cite{7b} and, more
recently, it was also pointed out that a mass term contribution of the
order $C m^2 H^2(t)$ (with $C>>1$) could smear the amplitude of the 
initial coherent oscillations definitely relaxing the problem \cite{5}.

Our purpose in this paper is not so ambitious and we will not
specifically address the moduli problem. We want
only to point out that the amplification of the gauge fields excited
by the coherent oscillations of the dilaton coupling
 should also be taken into account in the same
way as the scalar field fluctuations were shown \cite{2} to be
important in the standard Polonyi problem.
It is actually well known that in General Relativity the only fields
which can be amplified through the evolution of time-dependent
background geometry are the ones whose equations of motion are not
invariant under the Weyl rescaling of the metric tensor (such as, for
example, scalar and tensor inhomogeneities \cite{pert}) and since the evolution
equations of the Maxwell fields (and of the chiral fermions) are
invariant under the conformal rescaling of the metric background, it
was also correctly concluded that neither the electromagnetic
inhomogeneities nor the chiral fermions can be successfully amplified.
In string inspired backgrounds the situation  changes
and gauge fields fluctuations might grow thanks to the
dilaton coupling in the same way as the coupling to the background
curvature allows the parametric amplification of gravity waves in 
General Relativity.

The amplification of the electromagnetic fluctuations was discussed in
the context of the magnetic field generation \cite{11b}, but not
studied, so far, in the context of the
coherent oscillations of the dilaton field.
 
In the  low energy string
theory effective action \cite{5b} the evolution equations describing the
dynamics of the massless modes can be written either in the Einstein
frame (where the Einstein-Hilbert term and the dilaton field are
formally decoupled and the Planck scale is really a constant) or in
the String frame (where the coupling of the dilaton to the Einstein
Hilbert term is explicit and the typical scale of the theory is the
String scale [while the Planck scale is no longer a constants since it
evolves in time as $\lambda_P= e^{\frac{\phi}{2}} \lambda_S$]). As long
as the corrections in the string tension are not included the two
frames are equivalent (up to a conformal transformation and to a
ri-definition of the dilaton field). Moreover it was checked that the
amplitude of the excited inhomogeneities is exactly the same in both
the description even though the treatment in the String frame can be,
in principle, more cumbersome \cite{11c}. Since we are mainly concerned, in the
present discussion, with small curvature scales ($H< M_{P}$) and small
dilaton couplings ($e^{\phi/2}\ll 1$)we find useful to
work, from the very beginning, in the Einstein frame, where the low
energy effective action takes the form:
\begin{equation}
S= - \int d^4 x \sqrt{-g}(R -
\frac{1}{2}g^{\mu\nu}\partial_{\mu}\phi\partial_{\nu}\phi + V(\phi)+
\frac{1}{4} e^{-\phi} F_{\mu\nu}^{a} F^{a\mu\nu}+...)
\label{action}
\end{equation}
(we adopted ``natural'' units by setting $16\pi G=1$ and we neglected
the possible contribution of the antisymmetric tensor field which is
not directly relevant for the problem which we are discussing;
$F^{a}_{\mu\nu}$ is a generic gauge field which will be considered
abelian; $\phi= \Phi - \ln{V_6}= \ln{g^2}$ is connected with the four
dimensional gauge coupling [$\Phi$ is the 10-dimensional dilaton field
and $V_6$ is the volume of the 6-dimensional compact internal space] ).

The origin of the dilaton potential $V(\phi)$ is non perturbative and
related to supersymmetry breaking.
On theoretical ground, in critical superstring theory the dilaton
potential has to go zero as a double exponential as  $\phi\rightarrow
-\infty$ (weak coupling), namely $V(\phi)\sim \exp{(-c^2 \exp{(-\phi}))}$ with
$c^2$ a positive (model dependent) constant. On more physical ground
$V(\phi)$ is believed to have also one (or more) minima for some
finite value of the dilaton $\phi\sim \phi_{min}$. In general we can say
that after a phase of decreasing coupling driven either by the
dilaton's kinetic energy or by a radiation dominated background the
coherent oscillations start at a curvature scale $H\sim m$ \cite{10,11}.

The fluctuations of the gauge fields can then be amplified in two
different ways : either through the oscillating phase itself (for
Fourier modes $k \laq m$) or through the sudden change from the
(model-dependent) pre-oscillating epoch to the oscillating one.
In the first case the gauge fields are amplified in a resonant way
with a mechanism very similar to the ones recently introduced in the
studies of the so-called explosive
inflaton decay \cite{8,9} during the pre-heating dynamics 
\cite{8,9b,9c,9d}.
 In the second case the gauge fields are amplified in a
non-resonant way since the main effect will be given not by the
oscillations themselves but by the the change of  the background
evolution towards a phase effectively dominated by a non-relativistic
pressure-less fluid which ends up, ultimately, with the final dilaton
decay at a curvature scale $H_d\sim \Gamma \sim m^3/M_P^2$. 
The amplified electromagnetic inhomogeneities should not over-close the
universe and then the energy density of the fluctuations should be
always smaller than the energy density of the relativistic plasma. The
critical energy density bound applied on the amplified fluctuations
can then be translated into a bound on the parameters of the
background during the phase of the coherent oscillations, leading
 to possible bounds on the dilaton mass. 

The outlined picture of the dilaton relaxation might be unreliable for many
reasons. By studying the coherent oscillations only in
the vicinity of the minimum of the potential we will tacitely assume
that the potential is only quadratic. Detailed studies of the
preheating dynamics both in flat \cite{9d} and curved space \cite{curv}
showed that this approximation is quite misleading. The quadratic
potential is in fact, by itself, unstable under renormalization and a 
$\lambda\phi^4$ is needed from the one-loop divergences. In the
 case of the present paper, moreover, even higher powers are expected given the
 double exponential form of $V(\phi)$. Our results  will not
have the same degree of accuracy of  Ref. \cite{9d,curv} also because we do
not include the back-reaction effect of the produced fluctuations on
the dynamics of the background as carefully done in \cite{curv} in the
case of de Sitter space. 

The case which we are going to discuss is
also quite peculiar because of the exponential coupling of the dilaton
to the gauge fields which was not specifically considered in any of
the previous discussions of the preheating dynamics.
This limitations will be reflected in our
calculations and we will come back to this point later.

The plan of this paper is then the following.
Sec. II is devoted to the study of the resonant amplification of
electromagnetic fluctuations occurring for modes $k\gaq m$, while in
Sec. III we will turn our attention to the non-resonant case ($k\laq
m$) and we will show, within some very elementary example,
how the bounds on the amplified electromagnetic fluctuations can be
turned into bounds on the dilaton decay scale and then on the dilaton
mass.

\renewcommand{\theequation}{2.\arabic{equation}}
\setcounter{equation}{0}
\section{Resonant Amplification}

In a conformally flat metric of FRW type 
\begin{equation}
g_{\mu\nu} = diag(1, -a^2(t), -a^2(t), -a^2(t))
\label{metric}
\end{equation}
the background equations
derived from the action (\ref{action}) become
\begin{eqnarray}
H^2 &=& \frac{1}{12}( {\dot\phi}^2 + 2 V), ~~~
\dot{H} = - \frac{1}{4}{\dot\phi}^2
\nonumber\\
\ddot\phi &+& 3 H\dot\phi + \frac{\partial V}{\partial\phi}=0
\label{background}
\end{eqnarray} 
($H= (\ln{a})^{\cdot}$, $^{\cdot}= \partial/\partial t$).
For scales $H\laq H_{max} < M_{P}$ ( where $H_{max}$ is the maximal
curvature scale during the pre-oscillating phase ) the evolution of
the dilaton can be quite different depending upon the particular
model (see for instance \cite{12,10,11}). 
Our discussion does not rely on any special background during
the pre-oscillating phase but only on the existence of such a phase, 
occurring for $H>m$. We will only assume that the
curvature scale and the coupling constant ($g= e^{\frac{\phi}{2}}$)
are both decreasing in time ($\dot{H} <0$, $\dot{g}<0$).

For example in the post-big-bang evolution of the
String cosmological models the background (right after $H_{max}$) is
dominated by radiation ($H\simeq 1/2t$) and the dilaton
decreases according to the third equation in (\ref{background})
reaching a constant value before the onset of the coherently
oscillating regime, i. e. $\phi = \phi_0 + \phi_1
({H}/{H_{max}})^{1/2}$.
If the background is dominated, during the
pre-oscillating phase, by the dilaton's kinetic energy \cite{12,11}
($H\sim 1/3t$, $\rho_{\phi}\sim a^{-6}$) the dilaton decreases as
$\phi = \phi_0 + \phi_1 \ln{({H_{max}}/{H})}$.
The potential appearing in Eq. (\ref{background})
can be approximated (according to the considerations of the
Introduction) by \cite{gabriele}
\begin{equation}
V(\phi) = \frac{m^2}{2} (\phi - \phi_{min})^2 \exp{(-c^2\exp{(- \phi)})}~~.
\label{potential}
\end{equation}
In the  following discussion we will keep only the quadratic terms in
the potential. This approximation will allow our estimates but at the
same time we stress again that already the (flat space) solutions
of the theory including  quartic self-interactions 
might turn out qualitatetively and quantitatively
different giving rise to sizeable errors (up to $40$\%) in the number
of produced particles \cite{back}.

As soon as $ H \sim O(m)$ the dilaton starts oscillating with initial
amplitude $\phi_0\simeq O(1)$ in Planck units. Moreover, for
interactions of gravitational strength we also  expect
$\phi_{min}\simeq O(1)$ and, without fine-tuning, $|\phi_0 -
\phi_{min}| \simeq O(1)$.

In the oscillating regime a self-consistent solution of the
Eq. (\ref{background}) with the  potential (\ref{potential})
is :
\begin{equation}
\phi(t) = \phi_{min} + \frac{4}{\sqrt{3}}(\frac{\phi_0^{+} \cos{mt} +
\phi_0^{-} \sin{mt}}{mt}
+\frac{\phi_0^{+}\sin{mt} \cos^2{mt} +
\phi_0^{-}\cos{mt} \sin^2{mt}}{(mt)^2}) + O ( (\frac{H}{m})^3)
\label{dilsol}
\end{equation}
\begin{equation}
a(t) = a_0 (mt)^{2/3}[ 1 + \frac{\cos{mt}}{6 (mt)^2} 
- \frac{1}{24(mt)^2} + O ((\frac{H}{m})^3)]
\label{scale}
\end{equation}
(with $|\phi^{+}_0|^2 + |\phi^{-}_0|^2=a_0^2$).

The leading term (in $H/m$) of Eq. (\ref{scale}) turns out to be more
important in the case of the non-resonant amplification which rely on
the change in the background evolution among the pre-oscillating and
the oscillating phase. It is easy to transform the previous expressions
to conformal time $\eta$ ($a(\eta)d\eta=dt$) by keeping 
only the leading order behaviour in $H/m$ in Eq. (\ref{scale}).
We will use the conformal time in order to describe the non-resonant
amplification (where the expansion of the universe is mainly
concerned). On the other hand it is far more convenient to analyse the
resonant amplification in cosmic time where the solutions of the
background equations  can be
directly inserted into the gauge fields evolution equations, leading,
finally, to more tractable expressions.

From now on we will specialise our discussion to the case of abelian
fields ( but our discussion can be easily extended to the non-abelian
case provided $e^{\phi/2}\ll 1$).
The evolution equation for the Fourier modes of the abelian gauge
fields (derived from the action (\ref{action}) and in the gauge
$A_0=0$, $\nabla_{i}A^{i}=0$) are, for each of the two physical polarizations
\begin{equation}
{\cal A}_i(k)'' + [ k^2 - g(g^{-1})'']{\cal A}_i(k)=0,~~~
g(\phi)=e^{\frac{\phi}{2}}
\label{conformal}
\end{equation}
(where ${\cal A}_{i} = g(\phi)A_i$ and $' \equiv \partial/\partial\eta$).
It is  useful to rewrite the same equation in cosmic time 
\begin{equation}
\ddot{f}_{i}(\omega) + [ \omega^2 - G(t)] f_{i}(\omega)=0
\label{f}
\end{equation}
where $\omega = k/a$, $f_i = \sqrt{a} {\cal A}_i$ and 
\begin{equation}
G(t) =(\frac{g}{\sqrt{a}}) (\frac{\sqrt{a}}{g})^{\cdot\cdot}= 
\frac{7}{48} {\dot\phi}^2 - \frac{1}{2}\ddot{\phi} +
\frac{1}{6} \frac{\partial V}{\partial\phi} +\frac{V}{24}
\label{G}
\end{equation}
(the second exact equality follows directly by repeated use of the
background equations (\ref{background})). By inserting
Eq. (\ref{dilsol}) in Eq. (\ref{G}) and by keeping only the leading
terms in $H/m$ we obtain  that Eq. (\ref{f}) can be rewritten as 
\begin{equation}
\frac{d^2 f_{i}}{dz^2} + [\delta_{eff}(z) - 2
\epsilon_{eff}\cos{2z}]f_i=0,~~~z=\frac{mt}{2}
\label{mathieu}
\end{equation} 
where $\delta_{eff}= 4 (\frac{k}{m})^2 (z/2)^{-4/3}$ , $\epsilon_{eff}=
\epsilon z^{-1}$ ($\epsilon= 8/(3\sqrt{3})\phi^{+}_0 $ ; 
in order to write the previous equation we have chosen, for
simplicity, that $\phi_0^{-}=0$ in Eq.
(\ref{dilsol}) and we also assumed $\phi_{min}\simeq \phi^{+}_0$). 
If we focus our attention on the regime where $k>>m$
(i. e. $H>>m$) we could neglect, in the first approximation the
universe expansion and then $\delta_{eff}\simeq\delta = 4(k/m)^2
2^{4/3}$, $\epsilon_{eff}\simeq \epsilon$.
In this resonant approximation Eq. (\ref{mathieu}) becomes exactly the
well known Mathieu equation \cite{tricomi}. 
Notice that when $\epsilon=0$ and $\delta
= 2n +1$ ($n=0, 1,2...$) there is an exact solution of
Eq. (\ref{mathieu}) with period $2\pi$. If $\delta= 2n$ there is an
exact solution with period $\pi$. We can then say that the instability
boundaries in the $\delta-\epsilon$ plane will cross $\epsilon=0$ at
the points $\delta = (2n)^2$ and $\delta(2n+1)^2$ which are the
conditions for the parametric resonance to hold. This analysis can be
pursued very easily if $\epsilon \ll 1$ (i.e. $\phi^{+}_0 \ll 1$). 

A more
precise map of the boundaries of the unstable region can be obtained
in the $\delta-\epsilon$ plane to the desired order in $\epsilon$.
The analytic treatment shows that in this case the solutions of Eq. 
(\ref{mathieu}) are exponentially unstable but only in narrow
resonance bands (see, for instance, \cite{9,9c} and \cite{landau})
$|E_n| \simeq {n^{2n}
 2^{3-2n}}[(n-1)!]^{-1}\epsilon^n \delta^{1-n}$ with 
$\delta= n \mp \frac{E_n}{2}$ and $E_{n} <\delta$. 
Each resonance band with $n>1$ corresponds to
keeping terms $O(\epsilon^n)$ and then the width of each successive
resonance shrinks  but
near the center of each resonance band the solutions take the form
$f^{(n)}_{i}(z)= ({4\delta(k)})^{-1/4} \exp{(\pm\alpha^{(n)}_k z)}$
(with $\alpha_k^{(n)}$real and positive).

As it was extensively noticed in the context of the studies of the
preheating dynamics in the narrow resonance regime the amplification
of the inhomogeneities can only be very mild. 
Since in our case $\epsilon= 8 ( 3 \sqrt{3})^{-1}\sim 1.53$
 we might argue that, provided we
do not fine-tune $\phi_0^{+} \ll 1$ we are not in the narrow resonance
regime but in the broad resonance.  
We notice that the mode function $G(t)$ appearing in
Eq. (\ref{mathieu}) reduced simply to a damped cosinus only because we
kept only quadratic terms in the dilaton potential in the vicinity of
its minimum. 

As shown in \cite{9d} already the
inclusion of $\lambda\phi^4$ term (required by stability under
renormalisation) leads  to a Lame' equation where the cosinus
in the mode function is replaced by an elliptic cosinus. Needless to
say that the most accurate analysis is performed using the Lame'
equation.  In the present
case, however, the mode function would look even more complicated than
the one of the usual Lame' equation
(even if we  keep only quartic self-interactions in $V(\phi)$), because
of the exponential coupling of the dilaton to the gauge fields which
is very different from the usual couplings discussed in the context of
the preheating dynamics \cite{8,9,9d}. 

The previous (and following)
estimates based on Eq. (\ref{mathieu}) are then expected to be
drastically changed in a fully consistent non-linear calculation and
we only use it as a naive toy model but, none the less we find useful
to illustrate with simple examples the effect we are discussing.

If in the case $\epsilon \sim
1$ we can have a resonance (in flat space) this can be expected to be
shut off pretty  soon by the universe expansion. 
A simple Runge-Kutta method can be employed after
the second order differential equation (\ref{mathieu}) has been reduced to a
system of two first order in the variables $Q\equiv f_i$ and $P\equiv
f_i'$. We took $\phi_0^{+}=1$ and $\epsilon =8(3 \sqrt{3})^{-1}$ and we explored
different values of $k/m$ after having normalised $Q$ at the value of
the quantum mechanical fluctuations ($Q(z_0)=f(z_0) = 
({4\delta_{eff}(z_0)})^{-1/4}$), and $z_0=2$ 
[i.e. $t_0 =1/m$] is the initial integration time). 
The results illustrated in {\bf Fig. 1}   emphasize the importance of the
universe expansion. We plot $f_i(z)$ for different values of 
the parameters $k/m$ and $A= \phi^{+}_0$.

The growth of the solutions in flat space and the 
consequent particle production is seriously affected by the universe
expansion.
\begin{figure}
\vskip-0.5in
\vbox{\centerline{%
\hskip-4em\epsfxsize=.4\hsize\epsfbox{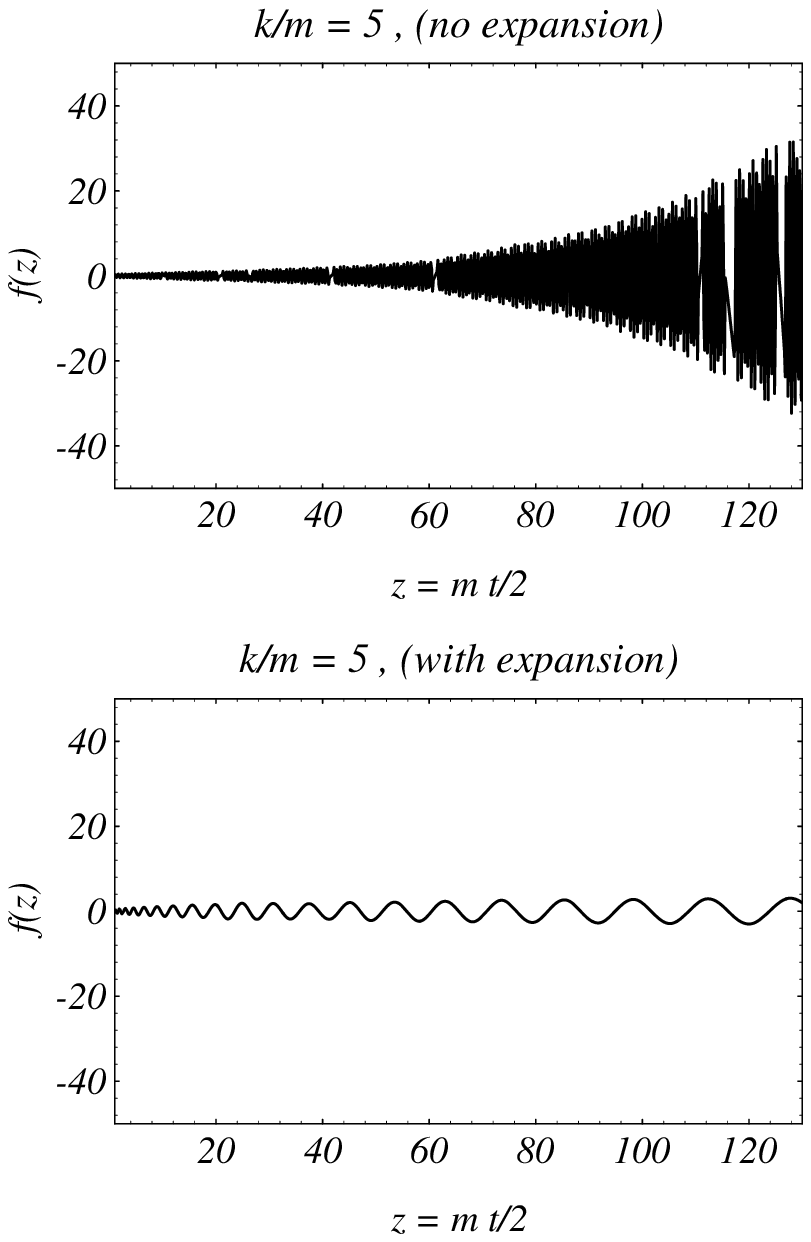}
\hskip-4em\epsfxsize=.4\hsize\epsfbox{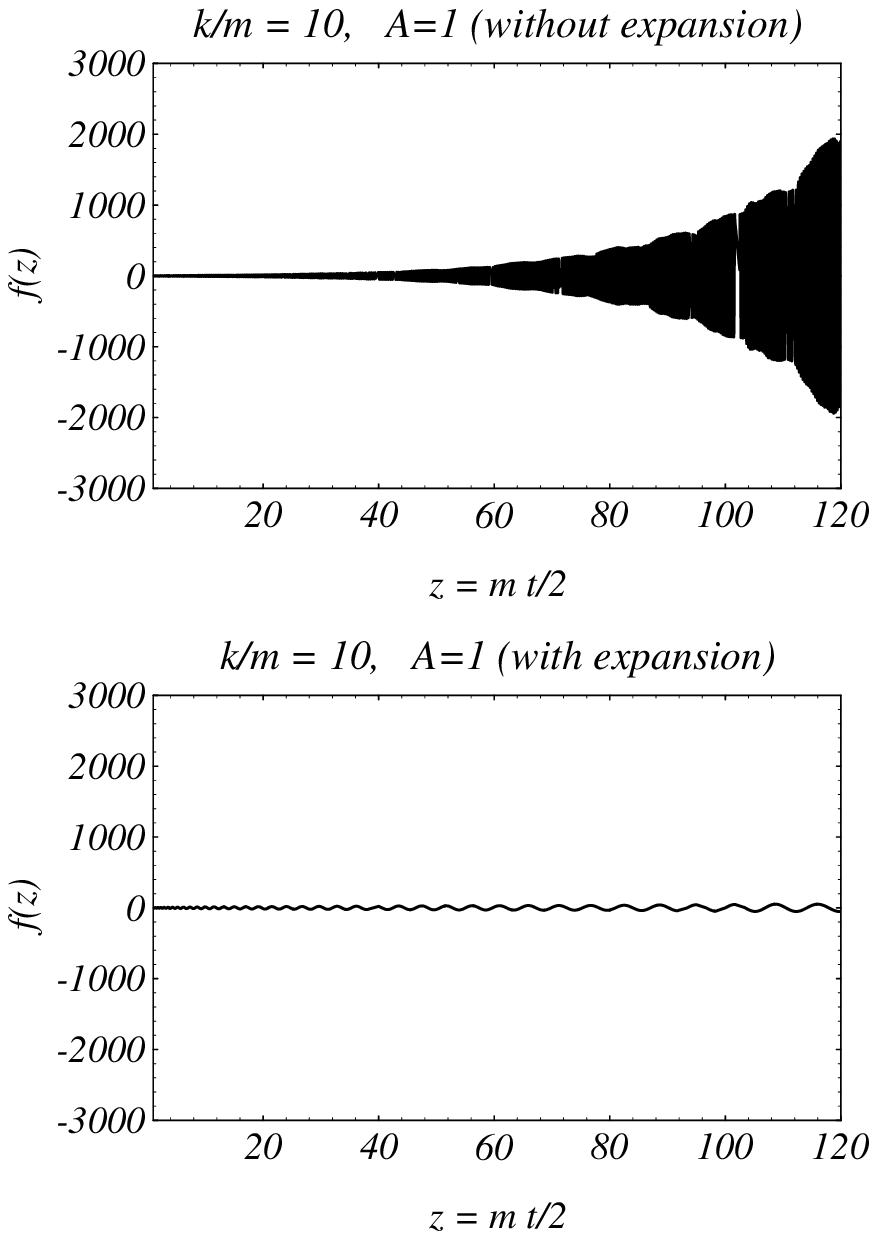}}
\vskip0.2in}
\caption{ We plot few  solutions of Eq. (2.11) for $A=\phi_0^{+}=1$
(i.e. $\epsilon =1.53$) in the resonant case ($k>m$). The effect of
the universe expansion on the flat space solutions is described.}
\label{fig-1}
\end{figure}
The solutions discussed in this section neglected completely the
problem of the back-reaction of the produced particles on the
evolution of the geometry. It was actually stressed in \cite{8,back}
that the effect of the produced inhomogeneities modifies the effective
equation of state from $p=0$ (in the regime of coherent oscillations)
to $p=\rho/3$ (when the produced fluctuations dominate). In our case
we can estimate that the back-reaction effect can be safely neglected
for times $\Delta t \sim 1/m (\Lambda + O(H/m))$ where $\Lambda$ is a
number which depends upon the particular model of dilaton
relaxation and upon the amplitude of the dilaton (in Planck units) at
the beginning of the oscilating phase.

\renewcommand{\theequation}{3.\arabic{equation}}
\setcounter{equation}{0}
\section{Non-resonant amplification}

We just showed that the amplification due to parametric resonance
seems quite small  and we now turn to the problem of non-resonant
amplification occurring for modes $k\laq m$. We will discuss in this
section a ``minimal''  toy model where the dilaton rolls down during a
(model dependent) pre-oscillating phase, settles to its minimum and
then decays at a curvature scale  determined by its decay
width $ H_d \sim \Gamma\sim m^3/M_P^2$ producing a radiation dominated
epoch. We are then led to study the
amplification of massless gauge fields in the following model :
\begin{equation}
g=e^{\phi/2}= (\frac{\eta}{\eta_{max}})^{\beta},~~
~a(\eta)\sim (\frac{\eta}{\eta_{max}})^{\alpha},~~~\eta<\eta_d,~~{\rm and}
~~a(\eta) \sim (\frac{\eta}{\eta_d}),~~~\eta>\eta_d
\label{model}
\end{equation}
(with $\beta<0$ and $\alpha<0$).

According to this simple model of background
evolution the effective potential $V(\eta)= g(g^{-1})''$ appearing in
Eq. (\ref{conformal}) will be :
\begin{equation}
V(\eta) = \frac{\beta(\beta+1)}{\eta^2},~~~~\eta<\eta_d~~~
{\rm and}~~V(\eta)=0,~~~~~\eta>\eta_d
\label{effpot}
\end{equation}

A particular solution of Eq. (\ref{conformal}) for ${\cal A}_{i}(k)$
can thus be written in terms of the first and second kind Hankel
functions $H^{(1)}$ and $H^{(2)}$ corresponding to free oscillating
modes in the $|\eta|\rightarrow \infty$  limit. 

The effective potential barrier leads to an amplification of the
initial fluctuations, ore equivalently, to particle production from
the vacuum. Starting with an incoming Fourier mode which is of
positive frequency with respect to the asymptotic state at the left of
the barrier ($\eta\rightarrow -\infty$), one has in general, for
$\eta\rightarrow\infty$ a linear combination of modes which are of
positive and negative frequency with respect to the new vacuum. 
The mixing coefficients $c_{\pm}$ define the Bogoliubov 
transformation connecting
the two vacua and determine the spectral distribution of
the amplified fluctuations. 

The general solution of Eq. (\ref{conformal}) in the two distinct
epochs defined in Eq. (\ref{model})-(\ref{effpot}) can be written for
each Fourier mode ${\cal A}_{i}(k)$
\begin{equation}
{\cal A}_i (k)= \frac{1}{\sqrt{k}}\sqrt{k\eta} H^{(2)}_{\mu}(k\eta),~~~~
\eta<\eta_{d},~~~{\rm and}~
{\cal A}_i (k)=\frac{1}{\sqrt{k}}(c_{+} e^{-ik\eta} + c_{-}
e^{ik\eta}) ~~~~\eta>\eta_d
\end{equation}
The Bogoliubov coefficients $c_{\pm}$ can be fixed by the two
conditions given by matching ${\cal A}_{i}$ and ${\cal A}_i'$ in
$\eta= \eta_d$. It is well known that the Bogoliubov coefficients 
determined in this ``sudden'' approximation can, in principle, lead to
ultraviolet divergences in the energy density of the amplified
fluctuations. The physical reason for this is that the modes with
frequency $k>1/\eta_d$ the sudden approximation cannot be rigourously
applied and the calculation should be performed with a smooth
effective potential barrier interpolating among the two  vacua. 

To smear
 the potential step will produce an exponential suppression
of $|c_{-}|$ for modes $k>1/\eta_{d}$. Then in the frequency band we
are interested in the amplification coefficient will be then given
by $|c_{-}| \simeq  (k\eta_d)^{-\frac{1}{2} - |\mu|}$ and 
the energy density of the produced massless gauge bosons $\rho$, per
 logarithmic interval of proper  frequency $d\ln{\omega}$,
 is obtained by summing over the two  polarisation states of the
 amplified waves:
\begin{equation}
\frac{d \rho}{d\ln\omega} = 2 N \omega^4 |c_{-}|^2
\label{spectra}
\end{equation} 
We included also the factor $N$ which counts the different massless
gauge fields appearing originally in the low energy string theory
effective action (\ref{action}) which can be, in principle, also quite
 large, of the order of $10^{3}-10^{4}$ (as it happens, for
example, in heterotic string theory  \cite{gabriele}).
In order to obtain the final expression of the spectral energy density of the
amplified fluctuations in critical units we divide Eq. (\ref{spectra}) 
by the radiation energy density produced by the dilaton decay 
($\rho_c= H_d^2 M_P^2 ({a_d}/{a})^4$)
and we obtain 
\begin{equation}
\Omega(\omega) =\frac{1}{\rho_c} \frac{d \rho}{d\ln\omega} \simeq
N (\frac{\omega}{\omega_d})^{3-2|\mu|} (\frac{H_d}{M_P})^2 
\end{equation}
The spectral energy density of the amplified inhomogeneities has to be
always $\Omega(\omega,t)<1$ for all the frequencies and for all the
times. If the energy spectrum is either flat or decreasing ($|\mu|\geq
3/2$) the critical density bound should be applied at the lowest
amplified frequency. If the  spectrum is increasing ($|\mu| <
3/2$) the bound
should be applied to the highest possible frequency which coincides,
in our case with $\omega\sim\omega_d$. Therefore the critical density
bound implies, respectively, 
\begin{equation}
\frac{m}{M_P}\laq N^{-\frac{1}{6}}
(\frac{\omega_i}{\omega_d})^{\frac{3-2|\mu|}{6}},~{\rm for}~~|\mu|\geq
\frac{3}{2}~~~{\rm and},~~~\frac{H_d}{M_P}\laq N^{-\frac{1}{2}},
~{\rm for}~~|\mu|< \frac{3}{2}
\label{condition}
\end{equation}
where $\omega_i$ is the infrared scale of the spectrum.
As an example we consider the case where during the pre-oscillating
phase the evolution is driven by the kinetic energy of the dilaton
starting from the scale $H_{max}<M_{P}$. In this case $\beta=
-\sqrt{3}$ and $|\mu| =(\sqrt{3}-1)/2$. The critical density bound
applied to the highest amplified frequency gives from
(\ref{condition}) in terms of the dilaton mass $m\laq N^{-1/6}
M_{P}$ (we used $H_d \simeq (m^2/M_P^3$). It is amusing that the critical density bound implies a constraint
for the dilaton mass which depends upon the number of the amplified
massless gauge fields. If $N\simeq 2 \times 10^{3}$  this bound basically implies
$m\laq 10^{-1} M_P$. Our example shows clearly that, in spite of the
simplified model of background evolution it is possible to use the
non-resonant amplification of massless gauge fields in order to
constrain the dilaton mass. Since our purpose, in this section, is
only illustrative, we only mention that it is possible to obtain more
complicated exclusion plots either by leaving $\mu$ as a free
parameter, or by adding to our ``minimal'' example the effect of
 the intermediate  phase
(occurring for $H_d<H<m$). We are anyway studying modes $k<m$
and the addition of this phase will only change the Bogoliubov coefficient
$|c_{-}|$ (and then the energy density of the amplified fluctuations)
 through the inclusion of an extra parameter (the duration
of the intermediate phase) which will make broader (in conformal time
units ) the ``under-barrier'' region in the effective potential
appearing in Eq. (\ref{effpot}) . 
In these cases further constraints relating the
dilaton decay scale to the other parameters of the background model
can be obtained. 

We discussed the resonant and non-resonant amplification of massless
gauge fields during a phase of dilaton oscillations and our
conclusions are that the
resonant amplification is a less  efficient mechanism unless the value
of the dilaton at the beginning of the oscillating phase is fine-tuned
to be orders of magnitude bigger than the Planck mass which appears to
be unreasonable for interactions of gravitational strength. 
The important caveat is, anyway, that the estimates we presented were 
done in some oversimplified model of dilaton relaxation and, for a
more reliable answer, a complete non-linear analysis would be needed.
The non-resonant amplification of massless gauge fields seem, already
in our example,  a more efficient
mechanism and can very likely lead to possible
constraints on the value of dilaton decay scale and on the value
of the dilaton mass.

\section*{Aknowledgments}
 
I am indebted to A. C. Davis and N. Turok for their  interest
in topics related to the present investigation and for the enjoyable
 atmosphere of DAMTP where part of the work was completed.
I would like to thank Andrei Linde for useful discussions which 
partially motivated this investigation and
 N. Sanchez for the stimulating environment created at the Erice
School of Astrophundamental topics where those discussions took
place. I am very grateful to M. Shaposhnikov for many important
 insights on the various roles of amplified electromagnetic inhomogeneities
and I  acknowledge M. Gasperini and G. Veneziano for
 valuable collaboration.


\begin{thebibliography}{99}

\bibitem{1} G. D. Coughlan, W. Fischler, E. W. Kolb, S. Raby and G. G.
Ross, Phys. Lett. {\bf B131}, 59 (1983).

\bibitem{2} A. S. Goncharov, A. D. Linde and M. I. Vysotsky,
Phys. Lett. {\bf B147}, 279, (1984).

\bibitem{3} G. German and G. G. Ross, Phys. Lett. {\bf B172}, 305 (1986).

\bibitem{4} J. Ellis, N. C. Tsamish and M. Voloshin, Phys. Lett. {\bf
B194}, 291 (1987).

\bibitem{5} J. Ellis, D. V. Nanopoulos and M. Quiros, Phys. Lett {\bf
B174}, 291 (1986).

\bibitem{6} A. Linde, Phys. Rev. D, {\bf 53} R4129 (1996).

\bibitem{12} T. Banks, M. Berkooz and P. Steinhardt, Phys. Rev. D {\bf
52}, 705 (1995).

\bibitem{5b} C. Lovelace, Phys. Lett. {\bf B135}, 75 (1984);E. S. Fradkin and
A. A. Tseytlin, Nucl. Phys. {\bf B261}, 1 (1985); C. G. Callan et al.,
Nucl. Phys. {\bf B262}, 593 (1985).

\bibitem{5c} E. Fischbach and C. Talmadge, Nature {\bf 356}, 207 (1992).

\bibitem{5d} T. Damour and A. M. Polyakov, Nucl. Phys. {\bf B423}, 532
(1994).

\bibitem{10} M. Gasperini, Phys. Lett. {\bf B327}, 214 (1994).

\bibitem{11} G. Veneziano, Phys. Lett. {\bf B265}, 287 (1991);
 M. Gasperini and G. Veneziano, Phys. Rev. D {\bf 50},  2519 (1994).

\bibitem{7b} D. H. Lyth and E. D. Stewart, Phys. Rev. Lett. 75, 201
(1995); Phys. Rev. D {\bf 53}, 1784 (1996).

\bibitem{pert}E. M. Lifschitz, Zh. Eksp. Teor. Fiz. 16,  587 (1946); 
 L. P. Grishchuk, Sov. Phys. JEPT {\bf 40}, 409 (1975); 
A. A. Starobinsky, Pis'ma Zh. Eksp. Teor. Fiz. {\bf 30}, 719 (1979).

\bibitem{11b} M. Gasperini, M. Giovannini and G. Veneziano, 
Phys. Rev. Lett. {\bf 75}, 3796 (1995); Phys. Rev. D {\bf 52}, 
6651 (1992).

\bibitem{8} L. Kofman, A. Linde and A. A. Starobinsky,
Phys. Rev. Lett. {\bf 73}, 3195 (1994); Phys. Rev. Lett. {\bf 76},
1011 (1996); E. W. Kolb, A. D. Linde and A. Riotto, Phys. Rev. Lett
{\bf 77}, 4290 (1996).

\bibitem{9} J. Traschen and R.  Brandenberger, Phys. Rev. D 42, 2491
(1990); Y. Shtanov, J. Traschen and R. Brandenberger, Phys. Rev. D
{\bf 51}, 5438 (1994).

\bibitem{9b} S.Yu. Khlebnikov and I.I. Tkachev 
Phys.Rev.Lett. {\bf 77}, 219 (1996); Phys.Lett. {\bf B390}, 80 (1997);
 PURD-TH-96-08, e-print Archive hep-ph/9610477 (unpublished).

\bibitem{9c} M. Yoshimura, Prog. Theor. Phys. {\bf 94}, 873 (1995);
H. Fujisaki, K. Kumekawa, M. Yamaguchi and M. Yoshimura, Phys. Rev. D
{\bf 53}, 6805 (1996).

\bibitem{9d} D. Boyanovsky, H. J. De Vega, R. Holman and
J. F. J. Salgado, Phys. Rev. D {\bf 54}, 7570 (1996).

\bibitem{curv} D. Boyanovsky, D. Cormier, H. J. de Vega and R. Holman,
Phys. Rev. D {\bf 55} (1997), in press.

\bibitem{gabriele} G. Veneziano, CERN-TH. 7502/94, Contribution to the
PASCOS'94 Conference, Syracuse N.Y., (May 1994).

\bibitem{back}D. Boyanovsky, H. de Vega, R. Holman, D.-S. Lee and
A. Singh, Phys. Rev. D {\bf 51}, 4419 (1995).

\bibitem{11c} M. Gasperini and M. Giovannini, Phys. Rev. D {\bf 47}, 
1519 (1993).

\bibitem{landau} L. D. Landau and E. Lifschitz, {\it Mechanics} page
80 ff. (Pergamon Press, Oxford, 1960).

\bibitem{tricomi} A. Erdelyi, W. Magnus, F. Obehettinger and F. R. Tricomi, 
{\it Higher Trascendental Functions} (Mc Graw-Hill, New York, 1953).

\end{thebibliography}
\end{document}